\documentclass{article}
\usepackage[margin=1.2in]{geometry}

\usepackage[dvipsnames]{xcolor}
\usepackage{graphicx}

\usepackage{bbm}
\usepackage{braket}
\usepackage[utf8]{inputenc} 
\usepackage[T1]{fontenc}    
\usepackage{hyperref}       
\usepackage{url}            
\usepackage{booktabs}       
\usepackage{amsfonts}  
\usepackage{amsmath}
\usepackage{amsthm}
\usepackage{nicefrac}       
\usepackage{microtype}      
\usepackage{xcolor}         
\usepackage{enumitem}
\usepackage{natbib}
\setlist[itemize]{align=parleft,left=0pt..1em}
\DeclareMathOperator{\Tr}{Tr}

\DeclareMathOperator*{\argmax}{arg\,max}

\theoremstyle{plain}
\newtheorem{theorem}{Theorem}[section]
\newtheorem{proposition}[theorem]{Proposition}

\theoremstyle{definition}

\theoremstyle{remark}

\title{Quantum Sequential Universal Hypothesis Testing \footnotetext[0]{The work of M. Zecchin and O. Simeone was supported by an Open Fellowship of the EPSRC (EP/W024101/1). O. Simeone was also supported by the EPSRC project (EP/X011852/1).}}
\usepackage{graphicx}
\usepackage{subcaption}
\author{%
	Matteo Zecchin\thanks{Centre for Intelligent Information Processing Systems,
		Department of Engineering, King’s College London, United Kingdom;
		\texttt{\{matteo.1.zecchin,osvaldo.simeone\}@kcl.ac.uk} },
	\quad
	Osvaldo Simeone\footnotemark[1],%
	\quad
	and Aaditya Ramdas\thanks{ Department of Statistics \& Data Science, Machine Learning Department, Carnegie Mellon University, Pittsburgh, USA;
		\texttt{aramdas@cmu.edu}}%
}
\date{}

\begin{document}

	\maketitle
	
	\begin{abstract}
		Quantum hypothesis testing (QHT) concerns the statistical inference of unknown quantum states. In the general setting of composite hypotheses, the goal of QHT is to determine whether an unknown quantum state belongs to one or another of two classes of states based on the measurement of a number of copies of the state. Prior art on QHT with composite hypotheses focused on a fixed-copy two-step protocol, with state estimation followed by an optimized joint measurement. However, this fixed-copy approach may be inefficient, using the same number of copies irrespective of the inherent difficulty of the testing task. To address these limitations, we introduce the quantum sequential universal test (QSUT), a novel framework for sequential QHT in the general case of composite hypotheses. QSUT builds on universal inference, and it alternates between adaptive local measurements aimed at exploring the hypothesis space and joint measurements optimized for maximal discrimination. QSUT is proven to rigorously control the type I error under minimal assumptions about the hypothesis structure. We present two practical instantiations of QSUT, one based on the Helstrom-Holevo test and one leveraging shallow variational quantum circuits. Empirical results across a range of composite QHT tasks demonstrate that QSUT consistently reduces copy complexity relative to state-of-the-art fixed-copy strategies.
	\end{abstract}

	\section{Introduction}
	\label{sec:intro}
	\subsection{Context and Motivation}
	\textit{Quantum hypothesis testing} (QHT) is a fundamental task in quantum information theory, with many applications in quantum technologies, including quantum sensing, quantum communication, and quantum cryptography \citep{wilde2013quantum, nielsen2010quantum,cheng2025invitation,Simeone_2025}. The objective of QHT is to determine which of multiple competing hypotheses describes a quantum state based on measurements performed on a number of copies of the state. In the most basic setting of binary QHT, there are two possible hypotheses, a null hypothesis and an alternative hypothesis. The hypotheses are said to be \emph{simple} when each corresponds to a single quantum state, or \emph{composite}  when one or both hypotheses correspond to sets of possible states.
	
	Testing composite hypotheses is significantly more challenging, as it must contend with the uncertainty about the state's identity within the true class. Prior work on QHT with general composite hypotheses has addressed a scenario involving a simple null and a composite alternative. In these settings, fixed-copy tests have been proposed that use a portion of the copies to estimate the state within the alternative hypothesis class, and then leverage this estimate to design a collective measurement on the remaining copies as the final test \citep{fujiki2025quantum}. 
	
	However, fixed-copy QHT methods cannot adapt the number of copies of the state to the intrinsic difficulty of the testing task, leading to potential inefficiencies. In fact, there may be states that are easier to assign to either hypothesis, while other states are more ambiguous, being close to states in both hypothesis classes. This has motivated the introduction of sequential QHT, in which the test can use a variable number of copies before reaching a decision \citep{martinez2021quantum,salek2022usefulness,li2022optimal}. 
	
	\textit{Sequential hypothesis testing} \citep{wald1947sequential,tartakovsky2014sequential,siegmund2013sequential} is a mature area of statistics that has been recently revitalized in the context of classical machine learning and artificial intelligence applications and modern experimentation pipelines in the information technology industry    \citep{ramdas2023game}. Sequential methods collect data adaptively, allowing the test to terminate as soon as sufficient evidence is available. For QHT, it was shown in \citep{martinez2021quantum} that, for \textit{simple} hypotheses, sequential QHT can reduce the average number of state copies as compared to fixed-copy settings, while preserving asymptotic error rate requirements.
	
	This work explores, for the first time, sequential QHT with composite hypotheses, introducing the \emph{quantum sequential universal test} (QSUT). In the presence of composite hypotheses, sequential tests can adaptively trade off resources used to estimate the state within the respective classes and to discriminate between the classes. To support this capacity, QSUT builds on the recently developed universal inference framework for classical, i.e., non-quantum, hypothesis testing \citep{wasserman2020universal}, alternating between adaptive local measurements aimed at exploring the hypothesis space and joint measurements optimized for maximal discrimination. QSUT provides a statistically principled method for controlling type I error for composite hypothesis testing problems under no assumptions on the composite hypothesis testing.
	
	\subsection{Background} Traditional approaches to QHT have primarily focused on the fixed-copy regime, with seminal work in this area including Helstrom’s theorem, characterizing the optimal measurement strategies for distinguishing simple hypotheses \citep{helstrom1969quantum} and asymptotic analyses of error exponents in both simple \citep{audenaert2007discriminating, hiai1991proper, ogawa2002strong} and composite settings \citep{hayashi2002optimal, bjelakovic2005quantum, berta2021composite, lami2025solution, hayashi2024generalized}. In the case of adaptive testing protocols, where measurements can be designed based on the outcomes of past measurements, it has been shown that adaptive strategies do not improve asymptotic testing performance, but they can provide gains in the finite-sample regime \citep{higgins2011multiple}. Practical instantiations of adaptive algorithms for quantum hypothesis testing, quantum channel discrimination, and quantum classification have been explored via the optimization of variational quantum circuits \citep{subramanian2024shallow,kardashin2022quantum,banchi2024statistical}.
	
	Fixed-copy practical testing strategies for composite QHT have been less investigated. Existing methods consider restricted settings, relying on group symmetries or the Gaussianity of quantum states \citep{hayashi2009group, kumagai2011quantum}, assuming simple null hypotheses \citep{fujiki2025quantum}, or focusing on one- or two-sample testing setups \citep{grootveld2025towards}. 
	
	Sequential testing methods offer an alternative paradigm. Unlike fixed-copy tests, which require the number of observations to be determined in advance, sequential tests collect data adaptively and stop once sufficient evidence has been collected to make a decision. In the non-quantum setting, sequential hypothesis testing is a mature field \citep{wald1947sequential,siegmund2013sequential}, and applications include two-sample testing  \citep{shekhar2023nonparametric}, change point detection \citep{tartakovsky2014sequential}, multi-armed bandits \citep{xu2021unified} and hyperparameter selection \citep{zecchinadaptive}, to name a few.
	
	In the quantum domain, sequential QHT has so far been addressed primarily from an information-theoretical perspective. Notable work in this area characterizes the copy complexity of sequential QHT for simple hypothesis setting in terms of the quantum relative entropy between the null and alternative hypotheses \citep{martinez2021quantum,salek2022usefulness}. The asymptotically optimal measurement strategy that minimizes the copy complexity under both hypotheses involves adaptively choosing, based on the current evidence, between the two measurements that maximize the forward and reverse quantum relative entropy between the two hypothesis states \citep{li2022optimal}. However, the optimization of the relative entropies is generally a complex problem \citep{berta2017variational}, and the strategy introduced in \citep{li2022optimal} does not offer finite-copy performance guarantees. Overall, valid non-asymptotic sequential QHT strategies remain unexplored for composite hypotheses. 
	
	\subsection{Contributions} In this work, we introduce QSUT, a novel framework for sequential QHT  in general settings with composite hypotheses. The key contributions of this work are:
	\begin{itemize}
		\item \textbf{Quantum sequential universal test (QSUT):} We introduce QSUT, a sequential QHT framework for composite hypotheses that supports adaptive measurement strategies and stopping rules, while providing rigorous type I error control. QSUT makes it possible to flexibly balance the need to collect information about the unknown states within the two classes with the optimization of maximally discriminating measurements based on the available information.
		
		\item \textbf{Specific implementations of QSUT:} We develop two specific instantiations of QSUT. The first, the \emph{adaptive learned Helstrom-Holevo test} (aLHT), adaptively refines a series of Helstrom-Holevo tests using information from past measurements. The second, the \emph{adaptive learner variational test} (aLVT) adapts measurements realized via shallow variational quantum circuits in an online fashion based on past observed outcomes.
		
		\item \textbf{Experimental validation:} We conduct an empirical study across different composite QHT tasks, demonstrating that the proposed QSUT reduces the average copy complexity compared to state-of-the-art QHT.
	\end{itemize}

	The remainder of the paper is organized as follows. In Section \ref{sec:background}, we introduce the composite QHT problem, review fixed-copy approaches to QHT, and define the key performance metrics used to evaluate these tests. Section \ref{sec:qst} presents the general framework for sequential QHT. In Section \ref{sec:seq_univ_quant_test}, we describe the proposed QSUT and its type I error control guarantees. This section also introduces practical measurement policies based on Helstrom-Holevo measurements and variational quantum circuits. In Section \ref{sec:exp}, we present a series of experiments demonstrating the advantages of QSUT over state-of-the-art fixed-copy QHT methods. Finally, Section \ref{sec:conclusion} concludes the paper and outlines potential directions for future research.

	\section{Background: Fixed-Copy Composite Hypothesis Testing}
	\label{sec:background}
	In this section, we describe the main notation adopted in this paper, and we review background material on fixed-copy quantum hypothesis testing.
	\subsection{Notation}
	Let $\mathcal{H}$ denote a $d$-dimensional Hilbert space. A quantum state is described by a density matrix $\rho$, which is a positive semidefinite operator on $\mathcal{H}$ with unit trace, i.e., $\Tr(\rho) = 1$. We denote the set of all density matrices operating over the space $\mathcal{H}$ as $\mathcal{D}(\mathcal{H})$. Given a state $\rho\in \mathcal{D}(\mathcal{H})$, we denote its $n$-fold tensor product as $\rho^{\otimes n}$. A positive operator-valued measure (POVM) $\mathcal{M}$ is defined by a finite set of outcomes $\mathcal{X}$ and by the corresponding positive semidefinite matrices $\{M_x\}_{x \in \mathcal{X}}$. According to Born's rule, the probability of observing the outcome $x \in \mathcal{X}$ when measuring the quantum state $\rho$ using the POVM $\mathcal{M}$ is given by\begin{align}
		\label{eq:born}
		\Pr\left[X=x \mid \rho, \mathcal{M}\right] = \Tr(\rho M_x).
	\end{align}
	We also use the notation $P(X|\rho,\mathcal{M})$ to denote the distribution implied by the probability  \eqref{eq:born}.
	\subsection{Fixed-Copy Composite Quantum Hypothesis Testing}
	\label{sec:comp_qht}
	In this work, we consider a general \textit{quantum hypothesis testing} (QHT) problem consisting of discriminating between two composite hypotheses, where the null and alternative hypotheses are defined by disjoint subsets of quantum states. Specifically, given a number of copies of a quantum system in an unknown state $\rho$, we wish to test the hypotheses
	\begin{align}
		H_0: \rho \in \mathcal{S}_0 \quad \text{vs.} \quad H_1: \rho \in \mathcal{S}_1,
	\end{align}
	where the null-hypothesis subset $\mathcal{S}_0\subseteq \mathcal{D}(\mathcal{H})$ and the alternative-hypothesis subset $\mathcal{S}_1\subseteq \mathcal{D}(\mathcal{H})$ are disjoint, i.e., $\mathcal{S}_0\cap \mathcal{S}_1=\emptyset$. 
	
	In the standard fixed-copy QHT setup, one assumes the availability of a number $n$ of copies of the unknown quantum state $\rho$. The test is defined by a binary POVM $\mathcal{M}^n = \{M^n_0, M^n_1 = I - M^n_0\}$, which is applied to the joint state $\rho^{\otimes n}$.  The measurement outcome $X\in\{0,1\}$ determines the testing decision $D\in\{0,1\}$. The null hypothesis $H_0$ is accepted if the outcome is $X=0$, yielding decision $D=0$, while it is rejected in favor of the alternative hypothesis $H_1$ if the outcome is $X=1$, i.e., $D=1$.  
	
	Two important figures of merit for a test defined by the POVM $\mathcal{M}^n$ are the type I and type II error probabilities. The type I error corresponds to the maximum probability of rejecting the hypothesis $H_0$ when it is true, i.e.,
	\begin{align}
		\label{eq:type_I_fixed}
		\alpha(\mathcal{M}^n)=\sup_{\rho \in \mathcal{S}_0}\Pr[X=1|\rho^{\otimes n},\mathcal{M}^n]=\sup_{\rho \in \mathcal{S}_0}\Tr(\rho^{\otimes n} M^n_1),
	\end{align}
	where the supremum is taken over all states in subset $\mathcal{S}_0$, while the type II error corresponds to the maximum probability of incorrectly accepting $H_0$ when the alternative hypothesis $H_1$ is true, i.e.,
	\begin{align}
		\label{eq:type_II_fixed}
		\beta(\mathcal{M}^n)=\sup_{\rho \in \mathcal{S}_1}\Pr[X=0|\rho^{\otimes n},\mathcal{M}^n]=\sup_{\rho \in \mathcal{S}_1}\Tr(\rho^{\otimes n} M^n_0).
	\end{align}
	In asymmetric QHT, for a given significance level $\epsilon_0 \in (0,1)$, the POVM $\mathcal{M}^n$ is designed to minimize the type II error $\beta(\mathcal{M}^n)$ subject to the type I error constraint $\alpha(\mathcal{M}^n) \leq \epsilon_0$. The resulting minimum type II error is denoted as $\beta_{\epsilon_0}(\mathcal{M}^n)$ \citep{berta2021composite}.

	\section{Sequential Quantum Hypothesis Testing} 
	
	\begin{figure}
		\centering
		\includegraphics[width=0.65\textwidth]{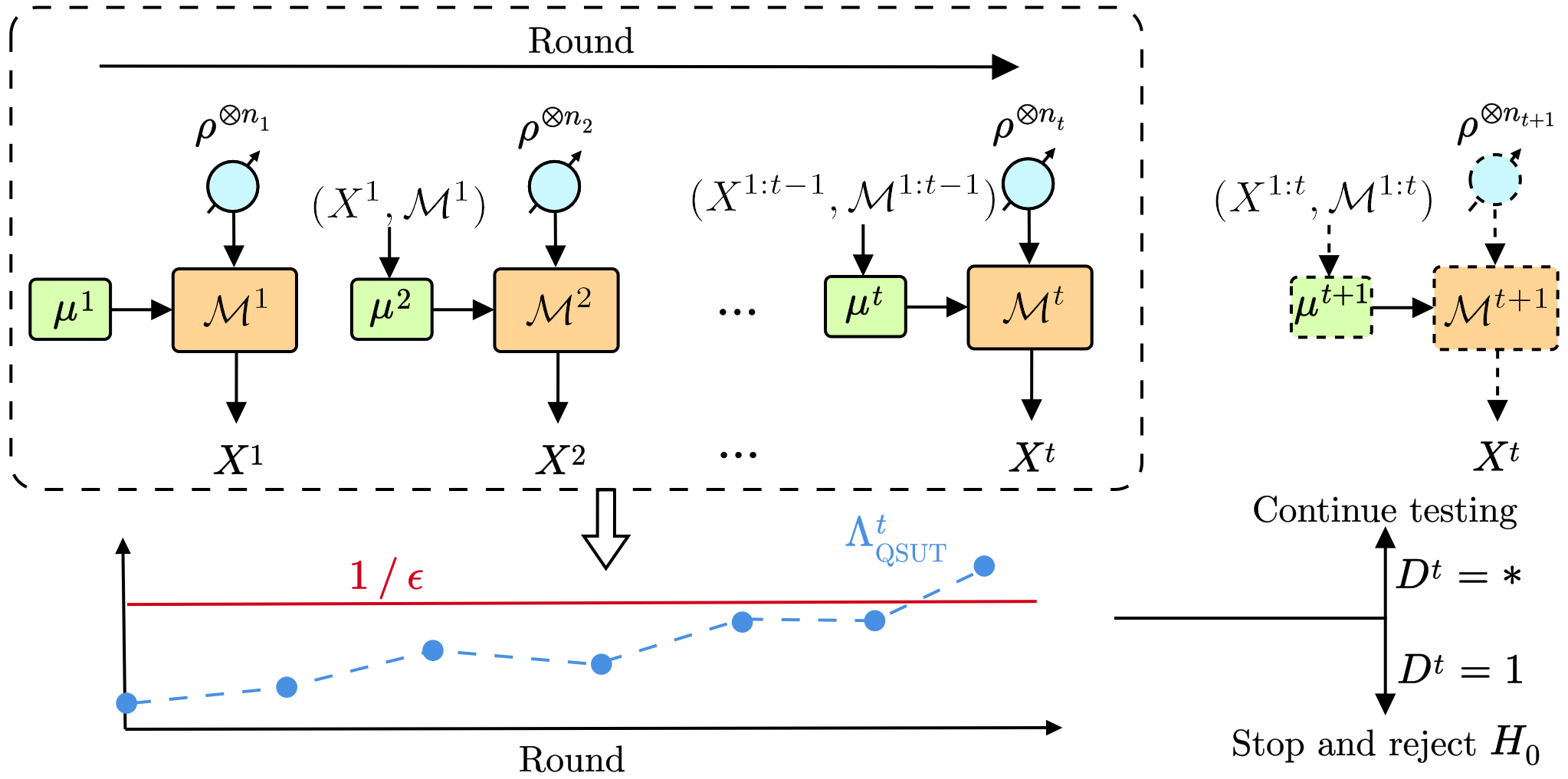}
		\caption{Illustration of the proposed quantum sequential  universal test (QSUT) framework. At each round $t \geq 1$, an arbitrary measurement policy $\mu^t$ selects the POVM $\mathcal{M}^t$. Then, it evaluates the non-anticipating sequential split likelihood ratio $\Lambda^t_{\rm QSUT}$ in \eqref{eq:e_process}, and it either halts or continues testing using the threshold rule \eqref{eq:q_decision_rule}.}
		\label{fig:squt}
	\end{figure}
	
	\label{sec:qst}

	As demonstrated in prior art on both non-quantum hypothesis testing \citep{wald1992sequential,robbins1972class} and QHT \citep{martinez2021quantum,li2022optimal}, a sequential formulation of testing strategies may offer copy efficiency gains. In fact, unlike the fixed-copy setting, in a sequential test, the number of copies $n$ consumed by the test is not determined in advance, making it possible to adapt the number of copies to the specific state $\rho$ being tested. Intuitively, this supports efficient QHT protocols in which ``easier'' states can be detected with fewer state copies, while ``harder'' states (close to both subsets $\mathcal{S}_0$ and  $\mathcal{S}_1$) require a larger number of copies. Prior work on sequential QHT focused, however, on simple hypotheses, while the goal of this work is to address the general case of composite hypotheses. 
	
	In \textit{sequential QHT}, testing proceeds in rounds as illustrated in Figure \ref{fig:squt}. At each round $t \geq 1$, based on the past measurement outcomes, the experimenter measures $n^t$ additional copies of the quantum state $\rho$ or terminates the testing procedure. If the test does not halt, a POVM $\mathcal{M}^t$ is used to measure the new state copies $\rho^{\otimes n^t}$, obtaining measurement outcome $X^t \sim P(X^t | \mathcal{M}^t, \rho^{\otimes n^t})$. We first focus on power-one sequential tests, i.e., one-sided tests, imposing only a constraint on the type I error \citep{robbins1972class}. This formulation is useful when one is interested in detecting an interesting event, represented by hypothesis $H_1$, as quickly as possible, while controlling the false detection probability. Accordingly, at each round $t$, the test outputs the decision 
	\begin{align}
		D^t=\begin{cases}
			* \quad \text{(continue testing),}\\
			1 \quad \text{(stop and reject $H_0$)}.
		\end{cases}
	\end{align}
	An extension to two-sided tests, which allow for the acceptance of the null hypothesis and thus for type II errors, is provided in Section \ref{sec:two_sided}.
	
	A sequential test is defined by a \textit{measurement policy} $\{\mu^t\}_{t \geq 1}$ and a \textit{decision rule} $\{\phi^t\}_{t \geq 1}$. Each function $\mu^t : (\mathcal{M}^{1:t-1}, X^{1:t-1}) \mapsto (n^t,\mathcal{M}^t)$ in the measurement policy maps the history of POVMs and measurement outcomes up to time $t-1$, respectively denoted as $\mathcal{M}^{1:t-1}$ and $X^{1:t-1}$, to the number of copies $n^t\geq1$ and to the corresponding  POVM $\mathcal{M}^t$ to be applied. Subsequently, the decision rule $\phi^t : (\mathcal{M}^{1:t}, X^{1:t}) \mapsto D^t$ maps the history of measurement settings and outcomes to the testing decision $D^t$. 
	
	A decision rule $\phi^t$ induces an observation-dependent stopping time $T$ at which a decision $D^t=1$ is produced, i.e.,
	\begin{align}
		\label{eq:tt}
		T = \inf\{t \in \mathbb{N} : D^t = 1\}.
	\end{align}
	We impose the condition that the type I error of this decision is no larger than a target value $\epsilon_0$ \citep{robbins1972class}, i.e.,
	\begin{align}
		\label{eq:seq_typeI_err}
		\sup_{\rho \in \mathcal{S}_0}	\Pr\nolimits[D^\tau = 1\mid \rho] \leq \epsilon_0.
	\end{align}
	In words, the probability of false alarm of any decision for hypothesis $H_1$ is upper bounded by $\epsilon_0$.
	
	The \textit{efficiency} of the test is measured by the number of copies of $\rho$ required to reject the null hypothesis when the alternative hypothesis $H_1$ is true. For any state $\rho\in\mathcal{S}_1$ this is quantified by the \emph{per-state copy complexity}
	\begin{align}
		\label{eq:q_compl}
		\bar{N}(\rho) = \mathbb{E}\left[\sum_{t=1}^T n^t \bigg|\rho\right], 
	\end{align}
	where we recall that $n^t$ denotes the number of copies measured at round $t$.   Also relevant is the \textit{per-state average termination time} \citep{wald1992sequential}, which counts the number of rounds prior to a decision $D^t=1$ as
	\begin{align}
		\bar{T}(\rho) = \mathbb{E} \left[T\mid \rho\right].
	\end{align}
	Note that both the copy complexity and the average termination time depend on the unknown ground-truth state $\rho$, as the latter determines the difficulty of the hypothesis testing problem.
	
	\section{Quantum Sequential Universal Test}
	
	\label{sec:seq_univ_quant_test}
	In this section, we introduce the \textit{quantum sequential universal test }(QSUT), an adaptive test for the sequential QHT problem with composite null and composite alternative hypotheses described in the previous section.  We first present the general form of the QSUT based on the universal inference framework \citep{wasserman2020universal} in Section \ref{sec:gen_qsut}, and then specialize the framework to: \textit{(i)} a sequential version of the fixed-copy test of \citep{fujiki2025quantum}, which is termed \textit{adaptive learned Helstrom-Holevo test} (aLHT); and  \textit{(ii)} to a sequential system with variational measurements \citep{liu2020variational,subramanian2024shallow}, which is referred to as \textit{adaptive learned variational test} (aLVT).
	
	\subsection{Quantum Sequential Universal Test: General Framework}
	\label{sec:gen_qsut}
	
	As illustrated in Figure \ref{fig:squt}, at each round $t$, the measurement policy $\mu^t$ of QSUT selects a POVM $\mathcal{M}^t$ based on the past measurements $\mathcal{M}^{1:t-1}$ and outcomes $X^{1:t-1}$  in any arbitrary way (see Section \ref{sec:qst}). Furthermore, the decision rule $\phi^t$ computes a statistic $\Lambda_{\rm QSUT}^t$, known as the \textit{non-anticipating sequential split likelihood ratio (SLR)},  based on the measurements and measurement outcomes, $\mathcal{M}^{1:t}$ and $X^{1:t}$, obtained so far. The decision $D_{\rm QSUT}^t\in\{*,1\}$ is then produced via a thresholding mechanism. As we detail next, the SLR $\Lambda_{\rm QSUT}^t$ evaluates the evidence available for the null hypothesis $H_0$ over the alternative hypothesis $H_1$.
	
	Using the  SLR $\Lambda_{\rm QSUT}^t$, the test rejects the null hypothesis $H_0$ if the ratio $\Lambda_{\rm QSUT}^t$ exceeds the threshold $1/\epsilon_0$; otherwise, the test continues  by performing additional measurements, i.e.,
	\begin{align}
		\label{eq:q_decision_rule}
		D_{\rm QSUT}^t=\phi(\Lambda_{\rm QSUT}^t )=\begin{cases}
			*, \quad \text{if} \ \Lambda_{\rm QSUT}^t< \frac{1}{\epsilon_0},\\
			1, \quad \text{otherwise}.
		\end{cases}
	\end{align}
	The QSUT decision rule \eqref{eq:q_decision_rule} induces the termination time \eqref{eq:tt}
	\begin{align}
		T_{\rm QSUT} = \inf\left\{t \in \mathbb{N} : \Lambda_{\rm QSUT}^t\geq \frac{1}{\epsilon_0} \right\}.
	\end{align}
	
	The  SLR $\Lambda_{\rm QSUT}^t$ is the ratio of the probabilities that the measurement outcomes $X^{1:t}$ obtained so far are consistent with either hypothesis. To evaluate these probabilities, at each round $t$, the QSUT first computes \textit{maximum likelihood estimates} (MLEs) under the null and alternative hypotheses as

	\begin{align}  
		\label{eq:mle_null}
		\hat{\rho}^t_0 \in \arg\max_{\rho \in \mathcal{S}_0} \left\{ \prod_{i=1}^t \Pr\left[X=X^i\Big|\rho^{\otimes n^i}, \mathcal{M}^i\right]=\prod_{i=1}^t \Tr\left(\rho^{\otimes n^i} M^i_{X^i}\right)\right\},
	\end{align}
	and 
	\begin{align}  
		\label{eq:mle_alternative}
		\hat{\rho}_1^{t-1} \in \arg\max_{\rho \in \mathcal{S}_1}\left\{ \prod_{i=1}^{t-1} \Pr\left[X=X^i\Big|\rho^{\otimes n^i},  \mathcal{M}^i\right]=\prod_{i=1}^{t-1} \Tr\left(\rho^{\otimes n^i} M^i_{X^i}\right)\right\},
	\end{align}
	respectively. Note that the MLE \eqref{eq:mle_alternative} for the alternative hypothesis $H_1$ at round $t$ corresponds to the state that maximizes the likelihood of the measurements $X^{1:t-1}$, excluding the latest observation $X^t$, while the MLE \eqref{eq:mle_null} for the null hypothesis $H_0$ at round $t$ considers all the available measurements $X^{1:t}$.
	
	With these estimates, the  SLR $\Lambda_{\rm QSUT}^t$ at time $t$ is obtained as the ratio between the likelihoods of the data $X^{1:t}$ under the two hypotheses, where the likelihood of each measurement $X^i $ is evaluated for $H_1$ under the MLE $\hat{\rho}_1^{i-1} $ and for $H_0$ under the MLE $\hat{\rho}_0^t $: 
	\begin{align}
		\label{eq:e_process}
		\Lambda_{\rm QSUT}^t= \prod_{i=1}^{t} \frac{\Tr\left((\hat{\rho}^{i-1}_1)^{\otimes n^i} M^i_{X^i} \right)}{\Tr\left((\hat{\rho}^t_0)^{\otimes n^i} M^i_{X^i} \right)}.
	\end{align}
	Intuitively, the  SLR \eqref{eq:e_process} monitors the evidence in favor of the alternative hypothesis $H_1$. In fact, when hypothesis $H_1$ is true, the numerator will tend to dominate the denominator, and the opposite is true when hypothesis $H_0$ holds. 
	
	Importantly, the SLR $\Lambda_{\rm QSUT}^t$ defines an \emph{e-process}, a nonnegative stochastic process that, under $H_0$, has an expected value bounded by 1 at any stopping time \citep{wasserman2020universal}. As shown in the next Section, this property makes it particularly suitable for sequential or anytime-valid hypothesis testing, providing rigorous control over Type I error under optional stopping.

	\subsection{Theoretical Guarantees}
	Thanks to the properties of universal inference \citep{wasserman2020universal}, QSUT offers a family of sequential tests satisfying the type I error requirement \eqref{eq:seq_typeI_err}.
	
	\begin{theorem}
		\label{th:anytime_validity}
		For a target type I error $\epsilon_0 \in (0,1)$ and any sequence of measurement policy $\{\mu^t\}_{t \geq 1}$, QSUT  satisfies the type I error guarantee
		\begin{align}
			\sup_{\rho\in \mathcal{S}_0}\Pr\left[	D_{\rm QSUT}^{T_{\rm QSUT}}=1\Big|\rho\right]\leq \epsilon_0.
		\end{align}
	\end{theorem}
	\begin{proof}
		See Appendix \ref{sec:proof_th1}.
	\end{proof}
	
	\subsection{Adaptive Learned Helstrom-Holevo Test}
	\begin{figure}
		\centering
		\includegraphics[width=0.6\textwidth]{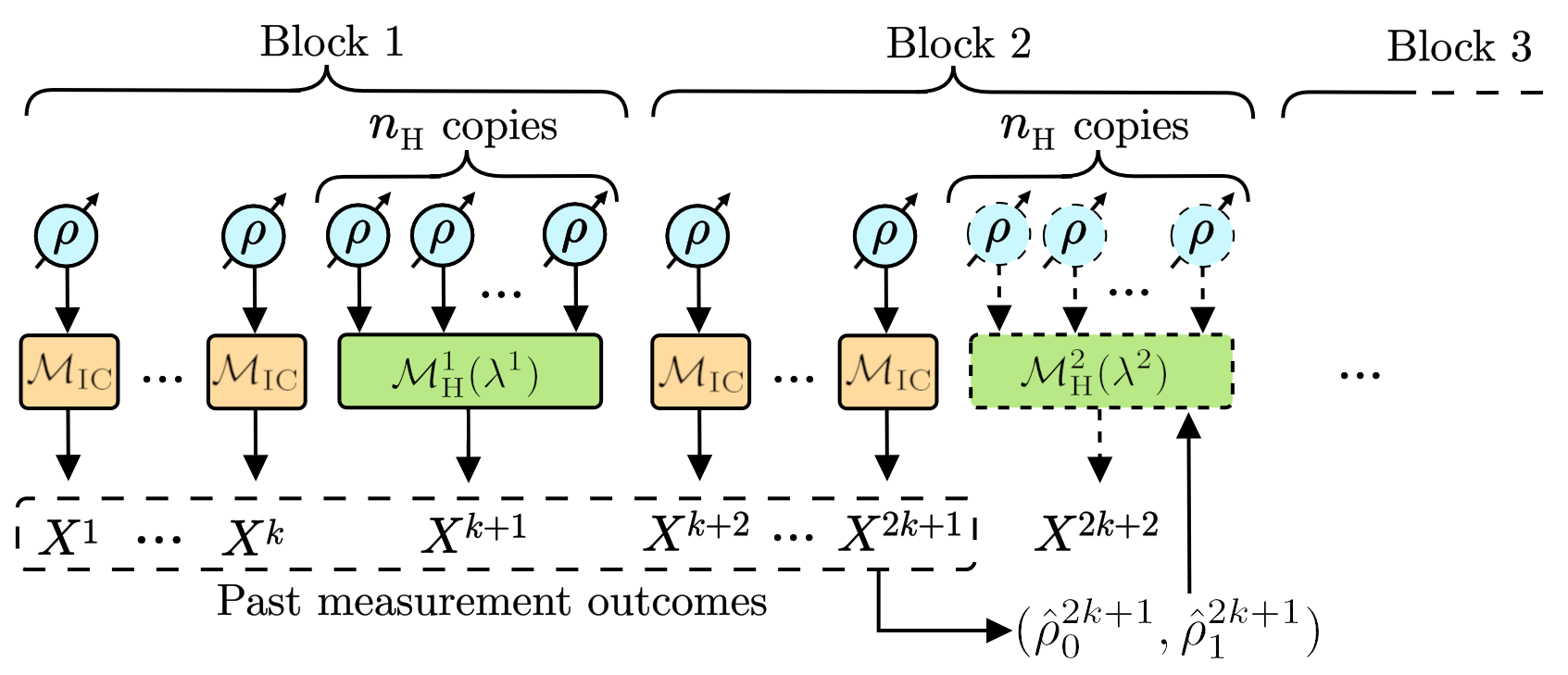}
		\caption{Illustration of the adaptive learned Helstrom-Holevo test (aLHT). The aLHT measurement policy follows a block structure. Each block comprises \(k+1\) rounds. During each of the first \(k\) rounds of a block, the test applies an informationally complete measurement \(\mathcal{M}_{\rm IC}\) to a single copy of the state \(\rho\). The final round of each block is used to compute a joint PVM \(\mathcal{M}^{b}_{\rm H}(\lambda^b)\) acting on \(n_{\rm H}\) copies of \(\rho\). The projectors of the PVM  \(\mathcal{M}^{b}_{\rm H}(\lambda^b)\) are determined by the projectors of the Helstrom-Holevo test computed using the null and alternative MLEs fitted on past measurement outcomes and a sensitivity parameter $\lambda^b$. The sensitivity parameter $\lambda^b$ can be fixed and predetermined, or adapted based on past measurement outcomes as in the adaptive learned Helstrom-Holevo test + (aLHT+).}
		\label{fig:sequential_fujiki}
	\end{figure}
	
	\label{sec:ALHT}
	QSUT supports the use of any measurement policy $\{\mu^t\}_{t\geq 1}$ for the selection of POVMs over rounds. In this subsection, we introduce a particular measurement policy that draws on the approach proposed in \citep{fujiki2025quantum} for the fixed-copy scenario with composite alternative $H_1:\rho\in\mathcal{S}_1$ and simple null $H_0:\rho=\rho_0$.
	The approach introduced in \citep{fujiki2025quantum}, termed here\textit{ learned Helstrom-Holevo test} (LHT), employs two types of measurements: the first is used to estimate the alternative state \( \hat{\rho}_1 \in\mathcal{S}_1\), while the second is a Helstrom-Holevo measurement \citep{helstrom1969quantum} designed based on state $\rho_0$ and $ \hat{\rho}_1$. The rationale underlying this design is that the Helstrom-Holevo measurement would be optimal when testing the simple hypotheses 	$H_0: \rho=\rho_0 \quad \text{vs.} \quad H_1: \rho=\hat{\rho}_1$  \citep{helstrom1969quantum}. Accordingly, the LHT replaces the state $\rho_1$ with the estimate $\hat{\rho}_1\in\mathcal{S}_1$ obtained using the first set of measurements.

	Building on this rationale, the proposed aLHT approach adopts measurement policy $\{\mu^t\}_{t \geq 1}$ that alternates between informationally complete measurements, used to estimate the unknown quantum states in subsets $\mathcal{S}_0$ and $\mathcal{S}_1$, and Helstrom-Holevo measurements, which are designed to discriminate between the estimated simple hypotheses.
	
	\textit{1) Adaptive learned Helstrom-Holevo test (aLHT):} As illustrated in Figure \ref{fig:sequential_fujiki}, the measurement policy $\mu^t$ implemented by aLHT follows a block structure defined by two positive integers, $k$ and $n_{\rm H}$. Each block $b$ consists of $k + 1$ rounds. In each of the first $k$ rounds of each block $b$, the unknown state is measured using an informationally complete POVM $\mathcal{M}_{\rm IC}$ \citep{renes2004symmetric}. Then, in the final round of each block $b$, a joint projective measurement (PVM) $\mathcal{M}^t=\mathcal{M}^b_{\rm H}(\lambda)$ is applied on $n_{\rm H}$ state copies following an Helstrom-Holevo measurement that is designed based on the null and alternative MLEs defined in \eqref{eq:mle_null} and \eqref{eq:mle_alternative}. Denote as $t_b=(k+1)b+k$ the number of rounds prior to the measurement in the $(k+1)$-th round of the $b$-th block. The corresponding MLEs in \eqref{eq:mle_null} and \eqref{eq:mle_alternative} are thus given by $\hat{\rho}^{t_b}_0$ and $\hat{\rho}^{t_b}_1$. 
	
	For a sensitivity parameter $\lambda\in(0,1)$, the Helstrom-Holevo measurement $\mathcal{M}^b_{\rm H}(\lambda)$ for the simple hypotheses $H_0:\rho=\hat{\rho}^{t_b}_0$ and $H_1:\rho=\hat{\rho}^{t_b}_1$ is given by two-outcome POVM
	$\mathcal{M}^b_{\rm H}(\lambda) = \left\{ M^b_0,I-M^b_0 \right\}$, where the measurement operator $M^b_0$ is given by the projector onto the positive eigenspace of the Hermitian matrix $(1 - \lambda)\, (\hat{\rho}^{t_b}_0)^{\otimes n_{\rm H}} - \lambda\, (\hat{\rho}^{t_b}_1)^{\otimes n_{\rm H}}$  \citep{helstrom1969quantum}. In asymmetric fixed-copy QHT, the parameter $\lambda$ quantifies the relative importance associated with type I and type II errors, effectively weighting the null and alternative hypotheses when constructing the optimal measurement. Larger values of $\lambda$ place greater weight on the alternative hypothesis $H_1$, and vice versa.

	\textit{2) Enhanced adaptive learned Helstrom-Holevo test (aLHT+):}
	In aLHT, the value of the sensitivity parameter $\lambda$ is fixed as a given hyperparameter. In the following, we propose aLHT+, a variant of aLHT in which the sensitivity parameter is optimized to reduce the copy complexity \eqref{eq:q_compl} of the sequential test under the alternative hypothesis $H_1$. 
	
	In non-quantum sequential hypothesis testing, a natural and widely adopted design criterion is the expected logarithmic increment of the test statistic under the alternative hypothesis $H_1$ \citep{shafer2021testing,grunwald2024safe,waudby2024estimating}. Following this principle, we propose designing the sequence of sensitivity parameters $\{\lambda^b\}_{b  \geq 1}$ for the measurements $\{\mathcal{M}^b_{\rm H}(\lambda^b)\}_{b \geq 1}$ used in each block $b\geq1$ by maximizing the logarithmic increment of the  SLR process $\{\Lambda^t_{\rm QSUT}\}_{t \geq 1}$, under the assumption that $\rho \in \mathcal{S}_1$. Specifically, we propose estimating this quantity by leveraging the MLEs $\hat\rho^{t_b}_1$ and $\hat\rho^{t_b}_0$, yielding the optimization
	\begin{align}
		\label{eq:approx_log_inc}
		\lambda^b=\argmax_{\lambda\in[0,1]}\mathbb{E}_{X^t\sim P(X|\hat\rho^{t_b}_1,\mathcal{M}_{\rm H}^b(\lambda))}\left[\log(\Tr((\hat\rho^{t_b}_1)^{\otimes n_{\rm H}} M^b_{X^t}))-\log(\Tr((\hat\rho^{t_b}_0)^{\otimes n_{\rm H}} M^b_{X^t}))\right],
	\end{align}  
	which can be efficiently solved using grid search methods.

	\subsection{Adaptive Learned Variational Test}
	\label{sec:AVT}
	\begin{figure}
		\centering
		\includegraphics[width=0.6\textwidth]{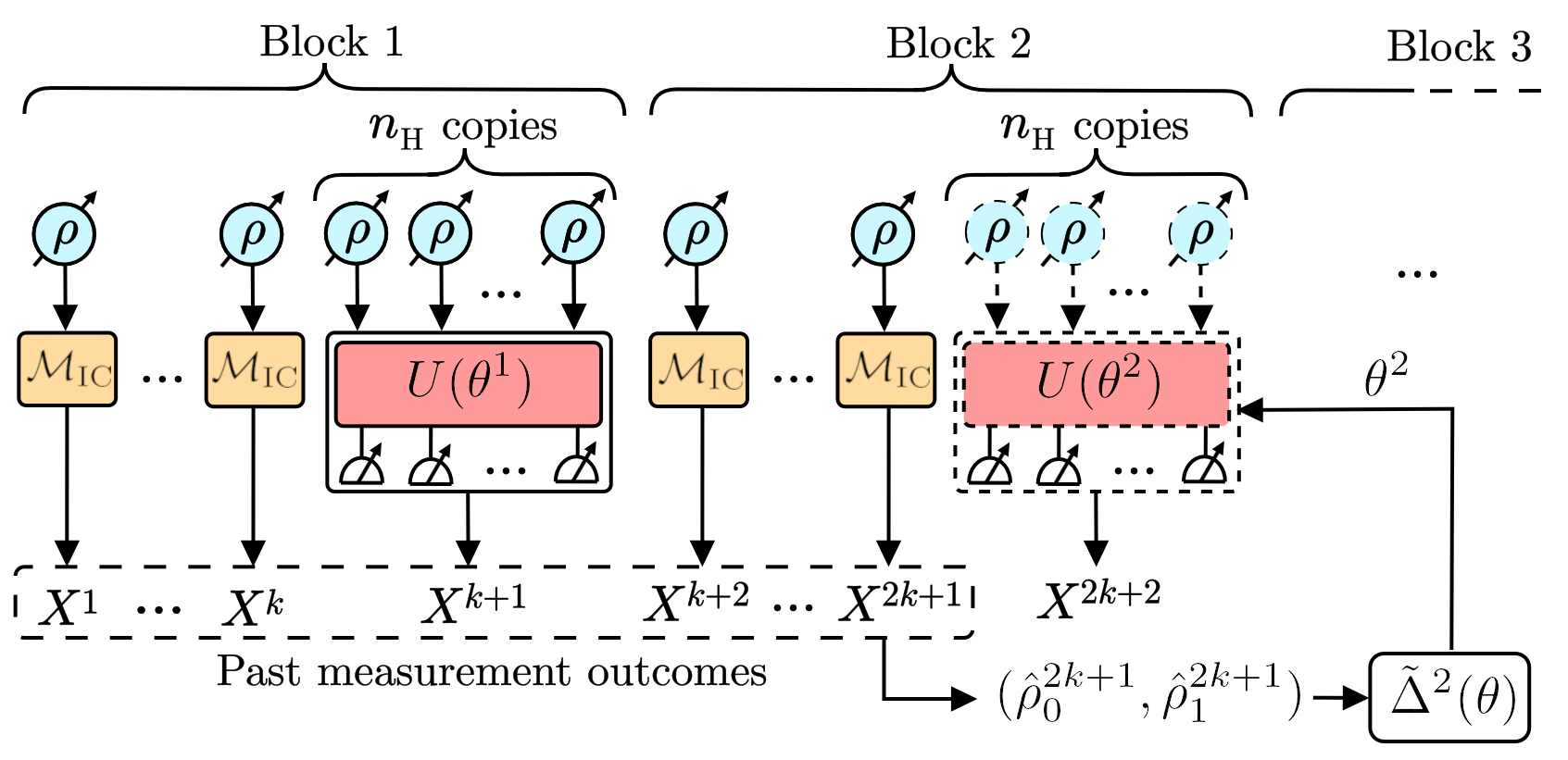}
		\caption{Illustration of the adaptive learned variational test (aLVT). The aLVT measurement policy follows a block structure. Each block comprises \(k+1\) rounds. During each of the first \(k\) rounds of a block, the test applies an informationally complete measurement \(\mathcal{M}_{\rm IC}\) to a single copy of the state \(\rho\). The final round of each block $b$ it corresponds to a measurement $\mathcal{M}^b_{\rm V}(\theta^b)$ which is realized applying a parameterized unitary quantum circuit $U(\theta^b)$ to the input state $\rho^{\otimes n_{\rm H}}$, followed by a standard computational basis measurement. The circuit parameters $\theta^b$ are optimized to reduce copy complexity of the test by maximizing an approximation of the expected logarithmic increment. }
		\label{fig:variational_fujiki}
	\end{figure}
	
	The Helstrom-Holevo measurement adopted by aLHT and aLHT+ is practically challenging to implement, as it generally requires quantum circuits of exponential depth in the number of copies being measured \citep{hwang2024quantum}. To address this issue, in this section we introduce aLVT, an alternative instantiation of QSUT that replaces the Helstrom-Holevo measurement with one implemented using shallow parameterized quantum circuits \citep{bharti2022noisy}.
	
	Specifically, aLVT follows the same block structure as aLTH and aLHT+ (see Figure \ref{fig:sequential_fujiki}), but it replaces the Helstrom-Holevo measurement $\mathcal{M}^b_{\rm H}(\lambda^b)$ with a variational measurement $\mathcal{M}^b_{\rm V}(\theta^b)$. The POVM $\mathcal{M}^b_{\rm V}(\theta^b)$ is realized by applying a parameterized unitary quantum circuit $U(\theta^b)$ to the input state $\rho^{\otimes n_{\rm H}}$, followed by standard local computational-basis measurements at each copy. This procedure specifies the POVM as
	\begin{align}
		\label{eq:variational_povm}
		\mathcal{M}^b_{\rm V}(\theta^b) = \left\{ M^b_x= U^\dagger(\theta^b)\Pi_x U(\theta^b) \right\}_{x \in \{0,1\}^{n_{\rm H}}},
	\end{align}
	where $\theta^b$ are the tunable parameters of the circuit at block $b$ and $\{\Pi_x=\ket{x}\bra{x}\}_{x \in \{0,1\}^{n_{\rm H}}}$ are projectors onto the computational basis.
	
	aLVT optimizes the parameters of the circuit $\theta^b$ to minimize the copy complexity of the sequential test. This is done by following the same design principle used in aLHT+ of maximizing the estimated expected logarithmic increment of the  SLR $\Lambda^t_{\rm QSUT}$ under the alternative hypothesis $H_1$ as in \eqref{eq:approx_log_inc}. Specifically, the optimization in \eqref{eq:approx_log_inc} is done here over the parameter $\theta^b$ of the POVM $\mathcal{M}^b_{\rm V}(\theta^b)$. This optimization can be carried out using standard gradient-based methods via the parameter-shift rule \citep{guerreschi2017practical,berta2017variational,simeone2022introduction}.
	\subsection{Extension to Two-sided Tests}
	\label{sec:two_sided}
	In this section, we introduce a two-sided variant of the QSUT test that, at each round~$t$, returns a ternary decision
	\begin{align}
		D_{\rm 2\text{-}QSUT}^t = 
		\begin{cases}
			* \quad \text{(continue testing),} \\
			1 \quad \text{(stop and reject $H_0$),} \\
			0 \quad \text{(stop and accept $H_0$).}
		\end{cases}
	\end{align}
	Unlike the power-one QSUT presented in Section~\ref{sec:gen_qsut}, the two-sided QSUT includes an additional outcome that allows the test to accept the null hypothesis $H_0$. This extension enables non-trivial control over both type~I and type~II errors. Specifically, in addition to satisfying the type~I error constraint given in~\eqref{eq:seq_typeI_err}, the two-sided QSUT is designed to ensure that, for a target error level $\epsilon_1$, the type~II error satisfies
	\begin{align}
		\label{eq:seq_typeII_err}
		\sup_{\rho \in \mathcal{S}_1} \Pr\nolimits[D^\tau = 0 \mid \rho] \leq \epsilon_1.
	\end{align}
	To guarantee that both types of errors are controlled, the two-sided QSUT maintains two parallel  SLR processes, denoted $\Lambda_{\rm QSUT,0}^t$ and $\Lambda_{\rm QSUT,1}^t$. The first process, $\Lambda_{\rm QSUT,0}^t$, is defined as in the power-one QSUT in ~\eqref{eq:e_process}, i.e., $\Lambda_{\rm QSUT,0}^t = \Lambda_{\rm QSUT}^t$, while the second process, $\Lambda_{\rm QSUT,1}^t$, has the same structure as $\Lambda_{\rm QSUT}^t$ but it reverses the roles of the null and alternative hypotheses. Concretely, at each time step~$t$, the two  SLR are defined as
	\begin{align}
		\label{eq:slrt_null}
		\Lambda_{\rm QSUT,0}^t= \prod_{i=1}^{t} \frac{\Tr\left((\hat{\rho}^{i-1}_1)^{\otimes n^i} M^i_{X^i} \right)}{\Tr\left((\hat{\rho}^t_0)^{\otimes n^i} M^i_{X^i} \right)},
	\end{align}
	and
	\begin{align}
		\label{eq:slrt_alt}
		\Lambda_{\rm QSUT,1}^t= \prod_{i=1}^{t} \frac{\Tr\left((\hat{\rho}^{i-1}_0)^{\otimes n^i} M^i_{X^i} \right)}{\Tr\left((\hat{\rho}^t_1)^{\otimes n^i} M^i_{X^i} \right)}.
	\end{align}
	For type I and type II error thresholds $(\epsilon_0, \epsilon_1)$, at each time $t$, the two-sided QSUT rejects the null hypothesis if the ratio $\Lambda_{\rm QSUT,0}^t$ exceeds $1/\epsilon_0$, accepts the null hypothesis if the ratio $\Lambda_{\rm QSUT,1}^t$ exceeds $1/\epsilon_1$, and otherwise continues testing if both ratios are below their respective thresholds. As shown in Appendix \ref{app:proof_prop}, the event where both $\Lambda_{\rm QSUT,0}^t$ and $\Lambda_{\rm QSUT,1}^t$ exceed their corresponding inverse thresholds is impossible if at least one of the thresholds $\epsilon_0$ or $\epsilon_1$ is smaller than 1. 
	
	The resulting decision rule of the two-sided QSUT is then given as
	\begin{align}
		D_{\rm 2\text{-}QSUT}^t=\phi(\Lambda_{\rm QSUT,0}^t,\Lambda_{\rm QSUT,1}^t)=\begin{cases}
			1, \quad \text{if} \ \Lambda_{\rm QSUT,0}^t\geq \frac{1}{\epsilon_0},\\
			0, \quad \text{if} \ \Lambda_{\rm QSUT,1}^t\geq \frac{1}{\epsilon_1},\\
			*, \quad \text{if $\Lambda_{\rm QSUT,0}^t< \frac{1}{\epsilon_0}$ and $\Lambda_{\rm QSUT,1}^t<\frac{1}{\epsilon_1}.$ }
		\end{cases}
	\end{align}
	Accordingly, the associated termination time is defined as
	\begin{align}
		T_{\rm 2\text{-}QSUT} = \inf\left\{t \in \mathbb{N} : \Lambda_{\rm QSUT,0}^t\geq \frac{1}{\epsilon_0} \text{ or } \Lambda_{\rm QSUT,1}^t\geq \frac{1}{\epsilon_1} \right\}.
	\end{align}
	
	Using arguments similar to those used for the power-one QSUT, the two-sided QSUT provides control over both type I and type II errors.
	\begin{theorem}
		\label{th:anytime_validity_two_sided}
		For a target type I error $\epsilon_0 \in (0,1)$, type II error $\epsilon_1 \in (0,1)$ and any sequence of measurement policy $\{\mu^t\}_{t \geq 1}$, the two-sided QSUT satisfies the type I error guarantees
		\begin{align}
			\sup_{\rho\in \mathcal{S}_0}\Pr\left[	D_{\rm 2\text{-}QSUT}^{T_{\rm 2\text{-}QSUT}}=1\Big|\rho\right]\leq \epsilon_0
		\end{align}
		and the type II error guarantee
		\begin{align}
			\sup_{\rho\in \mathcal{S}_1}\Pr\left[	D_{\rm 2\text{-}QSUT}^{T_{\rm 2\text{-}QSUT}}=0\Big|\rho\right]\leq \epsilon_1
		\end{align}
	\end{theorem}
	\begin{proof}
		The proof follows directly from Theorem \ref{th:anytime_validity} by noting that the two SLRs $\Lambda_{\rm QSUT,0}^t$ and $\Lambda_{\rm QSUT,1}^t$ are e-processes under hypotheses $H_0$ and $H_1$, respectively.
	\end{proof}
	\section{Experiments}
	\label{sec:exp}
	In this section, we present a series of experiments highlighting the advantages of the proposed QSUT test over fixed-copy QHT methods \citep{fujiki2025quantum}. Throughout, we focus on one-sided tests, imposing a type I error requirement $\epsilon_0$.
	
	\subsection{Setting}
	Following the setup in \citep{fujiki2025quantum}, we consider parametric QHT problems in which hypotheses are defined over a family of single-qubit quantum states given by
	\begin{align}
		\label{eq:state}
		\rho(\omega) = \frac{1}{2} \begin{bmatrix}
			1 + r_z \cos\omega & r_x \sin\omega \\
			r_x \sin\omega & 1 - r_z \cos\omega
		\end{bmatrix},
	\end{align}
	where $\omega \in \Omega=[0,2\pi)$ is an unknown parameter and the known constants $r_z$ and $r_x$ satisfy the constraint $r_z^2 \cos^2\omega + r_x^2 \sin^2\omega \leq 1$ to ensure validity of the density matrix. By \eqref{eq:state}, the parameters $r_z$ and $r_x$ determine the purity of the state, with larger values corresponding to higher purity. The hypotheses $H_0$ and $H_1$ are formulated in terms of admissible values of the parameter $\omega$, while parameters $r_z$ and $r_x$ are assumed to be known and fixed.
	
	We consider two distinct QHT scenarios. In the first, the null hypothesis is simple and given by $H_0:\omega = \omega_0$ for some known angle $\omega_0 \in \Omega$, while the alternative hypothesis is composite and defined as $H_1: \omega \in \Omega_1$ for some subset $\Omega_1 \subset \Omega \setminus \{\omega_0\}$. In the second scenario, both hypotheses are composite, i.e., $H_0: \omega \in \Omega_0$ and $H_1: \omega \in \Omega_1$, where the subsets $\Omega_0 \subset \Omega$ and $\Omega_1 \subset \Omega$ are disjoint.
	
	\subsection{Baselines}
	We compare the proposed QSUT against the following baseline procedures:
	
	\begin{itemize}
		\item \textbf{Learned Helstrom-Holevo test} (LHT) \citep{fujiki2025quantum} is a fixed-copy test for a simple null hypothesis. Given $n$ copies of the quantum state $\rho(\omega)$, the procedure operates in two phases. In the first phase, $m$ copies are measured using an informationally complete measurement, and the resulting outcomes $X^{1:m}$ are used to compute the MLE $\hat{\omega}_1^m$ of the alternative-hypothesis parameter $\omega_1$. In the second phase, the remaining $n - m$ copies are used to implement a Helstrom-Holevo measurement constructed for the test $H_0: \omega = \omega_0$ vs. $H_1: \omega = \hat{\omega}_1^m$, obtaining the decision $D_{\rm LHT}\in\{0,1\}$. The sensitivity parameter $\lambda \in (0,1)$ is tuned to satisfy the desired type I error constraint  \eqref{eq:type_I_fixed}.
		
		\item \textbf{Block learned Helstrom-Holevo test} (bLHT) is a direct extension of LHT introduced here to address the practical challenges of implementing a joint measurement on a large number of state copies. As in LHT, the first $m$ copies are used to estimate the alternative parameter $\hat{\omega}_1^m$. In the second phase, instead of performing a single Helstrom-Holevo measurement on all $n - m$ remaining copies, bLHT divides the $n-m$ into $b$ smaller blocks, where $n-m$ is assumed to be a multiple of $b$. Each block is independently tested using a Helstrom-Holevo measurement for test $H_0: \omega = \omega_0$ vs. $H_1: \omega = \hat{\omega}_1^m$, and the $b$ testing outcomes  $\{D^i_{\rm LHT}\}^b_{i=1}$ are combined using a majority vote scheme to obtain the final decision $D_{\rm bLHT}$. 
		
		\item \textbf{Learned variational test} (LVT) is a fixed-copy test introduced here as a benchmark applicable to both simple and composite hypotheses. It follows a similar two-phase approach as in LHT, but it replaces the Helstrom-Holevo measurement with a variational measurement realized via a shallow parameterized quantum circuit $U(\theta)$ followed by local measurements in the computational basis \citep{bharti2022noisy}. Given $n$ copies of the quantum state $\rho(\omega)$, the first $m$ are used to estimate the angle parameter $\hat{\omega}_1^m$ based on measurement in the computational basis, while the remaining $n - m$ are used for measurement with a variational circuit optimized to maximize the power of a likelihood ratio test (see Section \ref{sec:gen_qsut}). For a circuit parameter vector $\theta$, the variational POVM $\mathcal{M}_{\rm V}(\theta)$, defined as in $\eqref{eq:variational_povm}$, is used to compute the generalized likelihood ratio
		\begin{align}
			\label{eq:lr_lvt}
			\Lambda_{\rm LVT}(\theta) = \frac{\Tr(M_x \rho(\hat{\omega}_1^m))}{\sup_{\omega_0 \in \Omega_0} \Tr(M_x \rho(\omega_0))},
		\end{align}
		with the final decision rule produced by thresholding \eqref{eq:lr_lvt}. The threshold $\lambda$ is chosen to satisfy the type I error constraint \eqref{eq:type_I_fixed}, and the circuit parameters $\theta$ are optimized to minimize the type II error  \eqref{eq:type_II_fixed}, assuming the alternative state is $\rho(\omega) = \rho(\hat{\omega}_1^m)$.
		
		\item \textit{Block learned variational test} (bLVT) is further considered to scale LVT by replacing the single measurement with multiple measurements, whose outcomes are aggregated using a majority voting scheme, in a manner similar to bLHT.
	\end{itemize}

	\subsection{Simple Null and Composite Alternative}
	\begin{figure}
		\centering
		\begin{subfigure}[c]{0.425\textwidth}
			\centering
			\includegraphics[width=\textwidth]{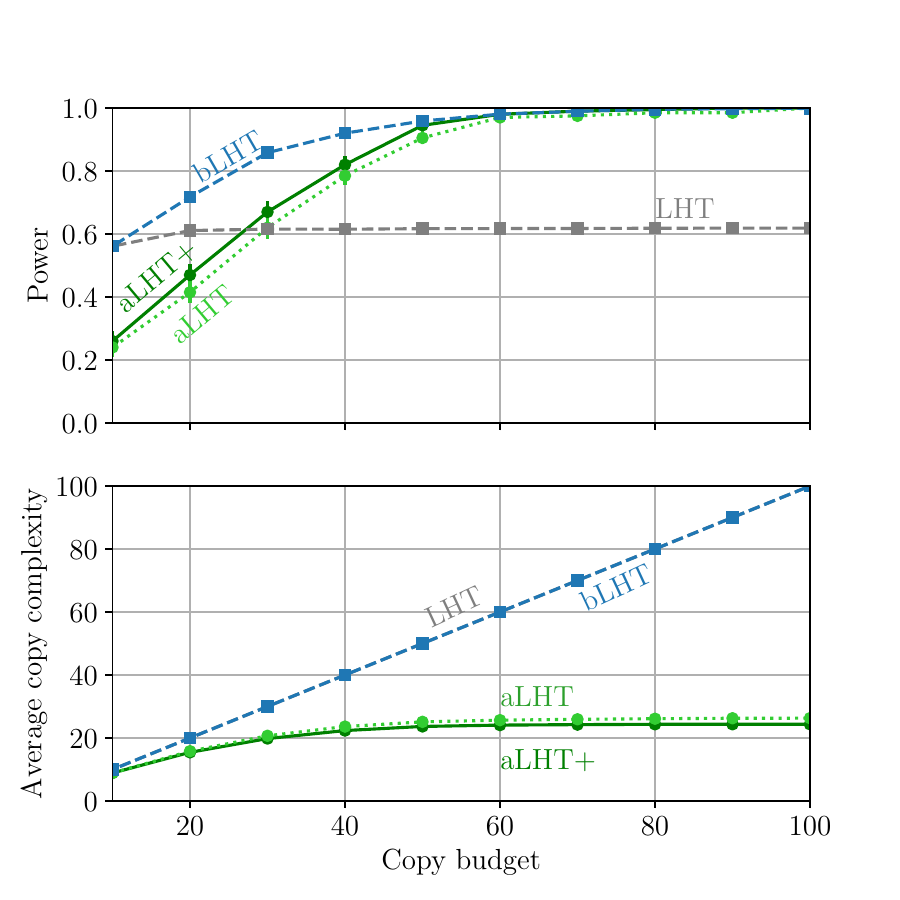}
			\label{fig:subfig1}
		\end{subfigure}
		\hfill
		\begin{subfigure}[c]{0.5\textwidth}
			\centering
			\includegraphics[width=\textwidth]{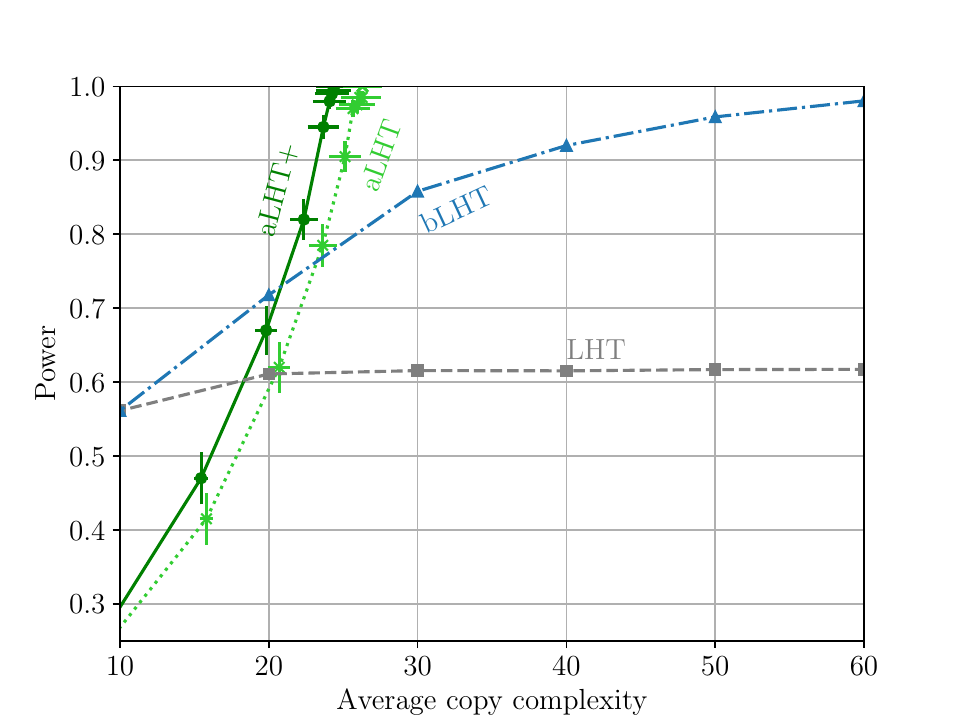}
			\label{fig:subfig2}
		\end{subfigure}
		\caption{Performance of the sequential tests aLHT+ and aLHT and the fixed-copy tests LHT and bLHT for QHT under a simple null hypothesis and a composite alternative hypothesis. In the left panel, we report the power and average copy complexity as functions of the copy budget, while in the right panel, the power is shown as a function of the average copy complexity. Due to their sequential and adaptive nature, aLHT+ and aLHT can terminate early when the available evidence is sufficiently conclusive, avoiding the need to consume the entire copy budget. This leads to higher power at fixed levels of copy complexity compared to LHT and bLHT, which are non-adaptive and always use the full budget.}
		\label{fig:simple_hypothesis}
	\end{figure}
	
	In this section, we consider QHT with a simple null hypothesis $H_0: \omega = 45^\circ$ and a composite alternative hypothesis $H_1: \omega \in (45^\circ, 180^\circ]$. For this QHT problem, we compare the performance of aLHT and aLHT+ against the fixed-copy schemes LHT \citep{fujiki2025quantum} and bLHT. 
	
	LHT uses measurements in the computational basis to estimate the alternative parameter $\hat{\omega}_1^m$, and performs a Helstrom-Holevo measurement over $n - m= 4$ copies. On the other hand, bLHT performs $b = \lfloor n / 10 \rfloor$ Helstrom-Holevo measurements on 4 copies each, using the remaining $m = n - 4b$ copies for the estimation of the parameter $\omega_1$. Note that while LHT always performs a single joint measurement on 4 copies regardless of the total copy budget $n$, bLHT scales the number of joint measurements with $n$.
	
	aLHT and aLHT+ are instantiated using blocks of size $k = 10$. The first $k - n_{\rm H} = 6$ copies  are measured in the computational basis, and a Helstrom-Holevo measurement $\mathcal{M}^b_{\rm H}(\lambda)$ is applied to the remaining $n_{\rm H} = 4$ copies. For aLHT, the sensitivity parameter $\lambda$ in $\mathcal{M}^b_{\rm H}(\lambda)$ is sampled uniformly at random from the interval $(0,1)$, whereas aLHT+ sets $\lambda$ by solving the optimization problem \eqref{eq:approx_log_inc} using a grid search method.

	In the left panel of Figure \ref{fig:simple_hypothesis}, we report the power and average copy complexity of the considered tests as functions of the copy budget $n$ for a ground-truth parameter $\omega=90^{\circ}$ and a target type I error level $\epsilon_0=0.05$. The power and copy complexity are averaged over 200 runs and, for a budget level $n$, the power of the test is evaluated as the fraction of runs in which the null hypothesis is rejected.  Importantly, for QSUT, the average copy complexity may be lower than the total budget $n$ since the test may reject $H_0$ early. In contrast, fixed-copy tests are non-adaptive and always consume all $n$ state copies, so that their average copy complexity equals the copy budget. The results in Figure \ref{fig:simple_hypothesis} show that for small values of $n$, all methods have similar average copy complexity, and fixed-copy tests achieve higher power. However, as $n$ increases, the power of aLHT and aLHT+ surpasses that of LHT, approaching that of bLHT, while requiring significantly fewer state copies on average.
	
	To better illustrate the efficiency gains from early stopping in QSUT, the right panel of Figure \ref{fig:simple_hypothesis} reports power versus average copy complexity. For low average complexity, fixed-copy tests yield higher power. However, once the average copy complexity exceeds 20, sequential tests become superior. Notably, LHT's power saturates around $0.6$ due to its fixed-size Helstrom-Holevo measurement, and bLHT requires nearly twice the number of state copies to reach a power of 0.9 compared to aLHT+.

	\subsection{Composite Null and Composite Alternative}
	\begin{figure}
		\centering
		\begin{subfigure}[c]{0.425\textwidth}
			\includegraphics[width=\textwidth]{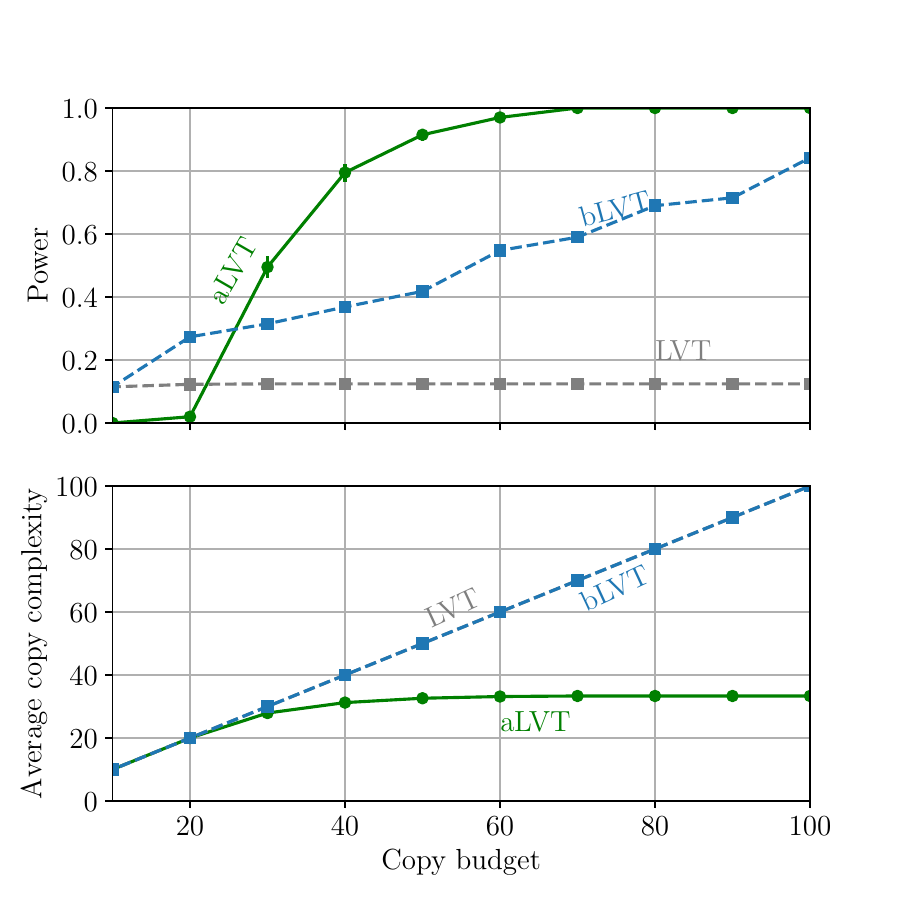}
		\end{subfigure}
		\hfill 
		\begin{subfigure}[c]{0.5\textwidth}
			\includegraphics[width=\textwidth]{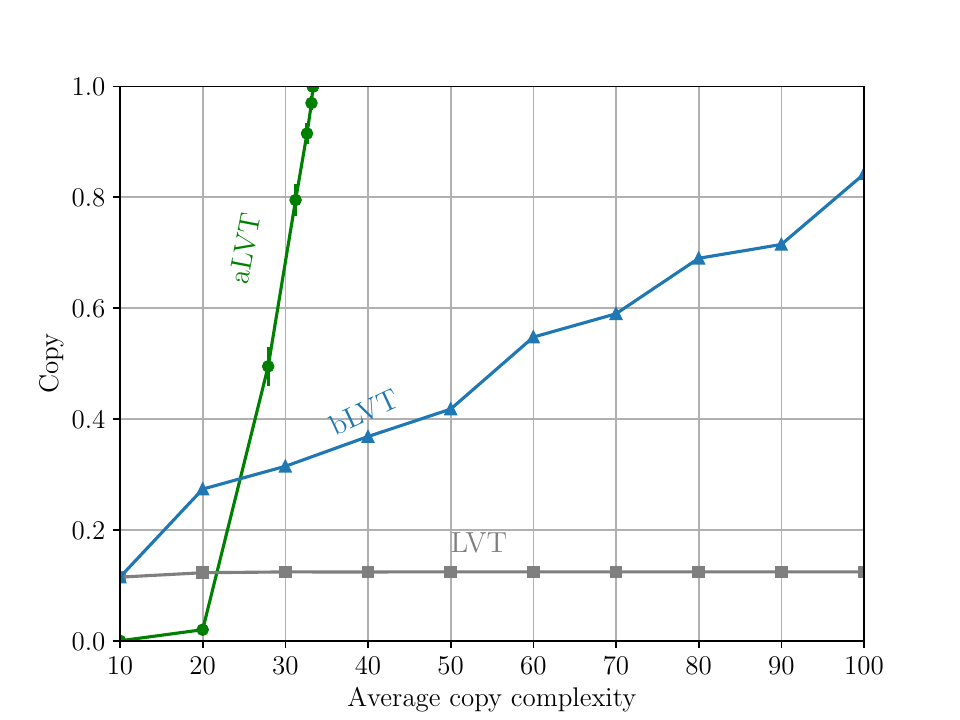}
		\end{subfigure}
		\caption{Performance of aLVT and the fixed-copy tests LVT and bLVT for QHT with a composite null hypothesis and a composite alternative hypothesis. In the left panel, we report the power and average copy complexity as functions of the copy budget, while in the right panel, the power is shown as a function of average copy complexity. Thanks to its ability to terminate testing early, aLVT can produce conclusive outcomes using only a fraction of the copy complexity required by LVT and bLVT, especially for large copy budgets.}
		\label{fig:composite_hypotheses}
	\end{figure} 
	
	We now consider a testing scenario in which both the alternative and null hypotheses are composite, defined as $H_0: \omega \in \{45^\circ, 135^\circ\}$ and $H_1: \omega \in (45^\circ, 135^\circ) \cup (135^\circ, 180^\circ)$. In this setting, we explore both sequential and fixed-copy approaches based on variational quantum circuits. More specifically, we compare the performance of aLVT introduced in Section \ref{sec:AVT} against the fixed-copy LVT and bLVT baselines. All tests employ variational measurements implemented using a hardware-efficient ansatz with a single layer consisting of Pauli $R_Y(\theta)$ rotations applied to each state copy, followed by a linear entangling layer of CNOT gates between adjacent state copies. The parameter $\theta$ of the circuit corresponds to the rotation angle of the $R_Y(\theta)$ gate and is shared across all copies.
	
	In both LVT and bLVT, the variational circuit acting on 4 copies is used to make the final decision. These schemes first apply computational-basis measurements to the first $m$ copies to estimate the alternative parameter $\omega_1$. In LVT, a single variational measurement is then applied to the remaining $n - m = 4$ copies. In contrast, bLVT applies the variational measurement $b = \lfloor n / 10 \rfloor$ times, and the outcomes are aggregated via majority voting. As such, in bLVT, the number of variational measurements scales with the copy budget $n$, whereas in LVT it remains fixed at one.
	
	aLVT is instantiated on blocks of $k = 10$ copies, where the first $k - n_{\rm V} = 6$ copies are measured in the computational basis, and the remaining $n_{\rm V} = 4$ copies are measured using the variational measurement $\mathcal{M}^b_{\rm V}(\theta^b)$ in \eqref{eq:variational_povm}. The parameters of this measurement are optimized via a grid search.

	In the left panel of Figure \ref{fig:composite_hypotheses}, we report the power and average copy complexity of the tests as functions of the copy budget $n$, for a ground-truth parameter $\omega = 90^\circ$ and significance level $\epsilon_0 = 0.05$. As in the previous section, all quantities are averaged over 200 runs. The results confirm the advantages of QSUT. While all methods exhibit similar power and complexity for a small copy budget $n$, as $n$ increases, aLVT not only surpasses LVT and bLVT in power but also achieves significantly lower copy complexity due to early stopping.
	
	This advantage is further highlighted in the right panel of Figure \ref{fig:composite_hypotheses}, which plots power versus average copy complexity. For small copy budgets, fixed-copy tests provide slightly higher power. However, as the copy complexity grows, aLVT delivers substantially greater power at a fraction of the average copy complexity. For example, aLVT achieves near-perfect detection of the alternative hypothesis with fewer than 40 copies on average, whereas LVT saturates at a power of 0.15, and bLVT requires nearly 100 copies to reach comparable performance.
	
	\section{Conclusion}
	\label{sec:conclusion}
	We have proposed QSUT, a new family of sequential QHT methods for the setting of composite hypotheses. QSUT leverages the universal inference framework introduced in \citep{wasserman2020universal} to define a sequential testing procedure with type I error control that supports adaptive measurement design aimed at reducing copy complexity. We presented two practical instantiations of QSUT, highlighting the flexibility of the framework to accommodate different implementation constraints, including scenarios in which complex joint measurements are infeasible. Through experiments on single-qubit hypothesis testing problems, we demonstrated that QSUT can substantially reduce copy complexity compared to state-of-the-art fixed-copy approaches, particularly when the underlying state is easy to distinguish.
	
	Our work opens several directions for future research, including the identification of the optimal measurement policy for QSUT that minimizes the average stopping time and the extension of the framework to two-sample hypothesis testing settings.

	\bibliography{ref}
	\bibliographystyle{plainnat}
	\appendix
	\section{Proof of Theorem \ref{th:anytime_validity}}
	\label{sec:proof_th1}
	For any $\rho\in\mathcal{S}_0$ define the stochastic process
	\begin{align}
		\label{eq:upper_bound_martingale}
		\bar{\Lambda}^t=\prod_{i=1}^{t} \frac{\Tr\left((\hat{\rho}^{i-1}_1)^{\otimes n^i} M^i_{X^i} \right)}{\Tr\left(\rho^{\otimes n^i} M^i_{X^i} \right)},
	\end{align}
	which corresponds to the  SLR $\Lambda_{\rm QSUT}^t$ in \eqref{eq:e_process} in which the MLE $\hat{\rho}^t_0$ is replaced by the ground-truth state $\rho$.
	By the definition of the MLE $\hat{\rho}^t_0$ in \eqref{eq:mle_null}, it follows that the  SLR $\Lambda^t_{\rm QSUT}$ is uniformly upper bounded by $\bar{\Lambda}^t$, i.e.,
	\begin{align}                                                                                                              
		\Lambda^t_{\rm QSUT}\leq \bar{\Lambda}^t.
	\end{align}                                                        
	Note that the stochastic process $\bar{\Lambda}^t$ is a supermartingale under $H_0$, in fact for any $ \rho \in \mathcal{S}_0$ it holds
	\begin{align}
		\mathbb{E}\left[\bar\Lambda^{t} \mid X^{1:t-1},\mathcal{M}^{1:t-1},\rho\right] &=\mathbb{E}\left[\frac{\Tr\left((\hat{\rho}^{t-1}_1)^{\otimes n^t} M^t_{X^t} \right)}{\Tr\left(\rho^{\otimes n^t} M^t_{X^t} \right)}\Bigg| \hat{\rho}^{t-1}_1,\rho\right]\prod^{t-1}_{i=1}\frac{\Tr\left((\hat{\rho}^{i-1}_1)^{\otimes n^i} M^i_{X^i} \right)}{\Tr\left(\rho^{\otimes n^i} M^i_{X^i} \right)}\nonumber\\
		&=\bar\Lambda^{t-1},
	\end{align}
	where the last equality follows from the fact that
	\begin{align}
		\mathbb{E}\left[\frac{\Tr\left((\hat{\rho}^{t-1}_1)^{\otimes n^t} M^t_{X^t} \right)}{\Tr\left(\rho^{\otimes n^t} M^t_{X^t} \right)}\Bigg|\hat{\rho}^{t-1}_1,\rho\right]=\sum_{x\in\mathcal{X}^t}\Tr\left((\hat{\rho}^t_1)^{\otimes n^t} M^t_{x} \right)=1.
	\end{align}
	Invoking Ville's inequality \citep{doob1939jean} applied to the supermartingale $\bar\Lambda^{t}$ we conclude that for the stopping time $T_{\rm QSUT}$ and any quantum state $\rho\in \mathcal{S}_0$ we have that
	\begin{align}
		\Pr\nolimits\left[	D_{\rm QSUT}^{T_{\rm QSUT}}=1\Big|\rho\right]=\Pr\left[\Lambda^{T_{\rm QSUT}}_{\rm QSUT}\geq\frac{1}{\epsilon_0}\Bigg|\rho\right]\leq \Pr\left[\bar\Lambda^{T_{\rm QSUT}}\geq\frac{1}{\epsilon_0}\bigg|\rho\right]\leq \epsilon_0\mathbb{E}\left[\bar\Lambda^1\big|\rho\right]=\epsilon_0.
	\end{align}

	\section{Impossibility of Simultaneous Crossing}
	\label{app:proof_prop}
	This appendix shows that the two  SLR processes $\Lambda_{\rm QSUT,0}^t$ and $\Lambda_{\rm QSUT,1}^t$ defining the two-sided QSUT test cannot simultaneously exceed their respective thresholds.
	\begin{proposition}
		\label{prop:double_crossing}
		For any $t \geq 1$ and error thresholds $\epsilon_0$ and $\epsilon_1$ such that $\min \{\epsilon_0, \epsilon_1\} < 1$, there is no sequence of POVMs $\mathcal{M}^{1:t}$ and measurement outcomes $X^{1:t}$ such that
		\begin{align}
			\left( \Lambda_{\rm QSUT,0}^t \geq \frac{1}{\epsilon_0} \right) \text{ and } \left( \Lambda_{\rm QSUT,1}^t \geq \frac{1}{\epsilon_1} \right).
		\end{align}
	\end{proposition}
	
	\begin{proof}
		We prove the statement by showing that if one process crosses its corresponding threshold, the other must remain below its threshold. Without loss of generality, assume the condition
		\begin{align}
			\Lambda_{\rm QSUT,0}^t \geq \frac{1}{\epsilon_0}.
		\end{align}
		From the  SLR definition \eqref{eq:slrt_alt}, we have
		\begin{align}
			\frac{\prod_{i=1}^{t} \Tr\left((\hat{\rho}^t_1)^{\otimes n^i} M^i_{X^i}\right)}{\epsilon_0} 
			\leq \prod_{i=1}^{t} \Tr\left((\hat{\rho}^{i-1}_0)^{\otimes n^i} M^i_{X^i}\right) 
			\leq \prod_{i=1}^{t} \Tr\left((\hat{\rho}^t_0)^{\otimes n^i} M^i_{X^i}\right),
		\end{align}
		where the last inequality follows from the definition of the MLE. Substituting this lower bound into the definition of $\Lambda_{\rm QSUT,1}^t$, we obtain the inequalities
		\begin{align}
			\Lambda_{\rm QSUT,1}^t 
			\leq \epsilon_0 \frac{\prod_{i=1}^{t} \Tr\left((\hat{\rho}^{i-1}_1)^{\otimes n^i} M^i_{X^i}\right)}{\prod_{i=1}^{t} \Tr\left((\hat{\rho}^t_1)^{\otimes n^i} M^i_{X^i}\right)} 
			\leq \epsilon_0,
		\end{align}
		where the last inequality again follows from the definition of the MLE. Since $\min\{\epsilon_0, \epsilon_1\} < 1$, we conclude that the implication
		\begin{align}
			\Lambda_{\rm QSUT,0}^t \geq \frac{1}{\epsilon_0} \implies \Lambda_{\rm QSUT,1}^t < \frac{1}{\epsilon_1},
		\end{align}
		holds, and a similar argument shows the reverse implication
		\begin{align}
			\Lambda_{\rm QSUT,1}^t \geq \frac{1}{\epsilon_1} \implies \Lambda_{\rm QSUT,0}^t < \frac{1}{\epsilon_0}.
		\end{align}
	\end{proof}

\end{document}